\documentclass[12pt]{article}
\usepackage{amsfonts}

\begin{document}
\title{Non Abelian structures and the geometric phase of entangled qudits}
\author{~L.~E.~Oxman and A.~Z. ~Khoury 
\\ \\ Instituto de F\'{\i}sica, Universidade Federal Fluminense,\\
Campus da Praia Vermelha, Niter\'oi, 24210-340, RJ, Brazil.
}
\date{\today} 
\maketitle 

\begin{abstract} 
 
In this work, we address some important topological and alge-
braic aspects of two-qudit states evolving under local unitary opera-
tions. The projective invariant subspaces and evolutions are connected
with the common elements characterizing the $\mathfrak{su}(d)$ Lie algebra and
their representations. In particular, the roots and weights turn out to
be natural quantities to parametrize cyclic evolutions and fractional
phases. This framework is then used to recast the coset contribution
to the geometric phase in a form that generalizes the usual
monopole-like formula for a single qubit.

\end{abstract}

%{\bf Keywords}: \\


\section{Introduction}

Entanglement is an essential component in quantum information protocols. The ability to operate entangled states without destroying their main features is often the central task in experimental implementations. Pure state entanglement can be measured by the concurrence \cite{1}, which is insensitive to local unitary operations on the individual subsystems. Under these evolutions, the geometric phase acquired by maximally entangled pairs of qubits has been predicted to occur in discrete steps \cite{2, 3, 4}. This discussion has been recently extended to multiple qubits \cite{5}. Phase steps, where a factor $e^{i\pi}$ is introduced, have been experimentally demonstrated with qubits encoded on spin-orbit laser modes \cite{6}, nuclear spins \cite{7} and entangled photon pairs \cite{8}. 

In ref. \cite{9}, based on the kinematic approach developed by Mukunda and Simon \cite{10, 11}, we investigated entangled qudits under unitary local operations, identifying some geometrical and topological aspects. In particular, the geometric phase was calculated in terms of the concurrence, and fractional
phases in cyclic evolutions were identified and analyzed. The extension to pairs of qudits with different dimensions was done in ref. \cite{12}, where the overlap of the evolving and initial state was illustrated with numerical examples in two-qutrit and qubit-qutrit systems. 

Experimental setups for the observation of fractional phases for entangled qudits \cite{13} and multiple qubits \cite{14} have already been proposed. Quantum gates based on geometric phases have been studied in the literature as a robust means for quantum computation \cite{15, 16}. In addition, fractional phases have been conjectured as a possible resource for fault tolerant quantum computation, though associated with a different physical situation. Namely, the fractional statistics due to the multiply connected nature of the configuration
space of anyons \cite{17}. Because of the various experimental and theoretical contexts involved, it is worth seeking for a thorough understanding of entangled qudit pairs operated by local unitary evolutions. 

In this work, we provide further insight into the different 
mathematical aspects involved. Initially, we shall obtain the fundamental homotopy group for the projective space of separable states and that for general rank-d states. Next, we will show how the $\mathfrak{su}(d)$ Lie algebra structure provides the appropriate tools to characterize two-qudit states, fractional and geometric phases. For example, evolutions containing a Cartan factor along the weights of $SU(d)$ representations are those generating fractional phases. Moreover, by using a time-dependent Lie algebra basis, we will show how to write the coset contribution to the geometric phase in terms of local Cartan elements $n_q$, $q=1,\dots, d-1$, projected along the fundamental weights of $\mathfrak{su}(d)$. These contributions will correspond to a generalization of the monopole-like Berry phase for a single qubit, where the phase can be expressed 
as the flux of a topological charge density on $S^2$, for an $S^2 \to \hat{n}\in  S^2$ mapping. The mathematics involved turns out to be that needed to discuss center vortices \cite{18}-\cite{22} and non Abelian monopoles \cite{23} in Yang-Mills-Higgs models with $SU(d) \to Z(d)$ spontaneous symmetry breaking.

In section 2, we review some general properties of invariant projective
subspaces and compute the fundamental homotopy groups for separable and
rank-d pure states. In section 3, we relate local evolutions and non Abelian connections, defined in terms of a local Lie algebra basis. In that section, we also show what are the possible evolutions leading to fractional phases. Section 4 is devoted to identify the Mukunda-Simon geometric phase as a sum weighted by d invariants under local evolutions, as well as by the weights of the fundamental $SU(d)$ representation. This phase is then carefully worked out to recast the coset sector as a superposition of monopole-like contributions. Finally, in section 5 we present our conclusions.

\section{The topology of invariant subspaces}

In Quantum Mechanics, an important concept is that of the projective space of states, which is essentially a topological space such that different points represent physically distinct quantum states. This space can be obtained by considering the equivalence relation between (normalized) kets,
\begin{equation}
|\psi\rangle \sim |\psi'\rangle \makebox[.5in]{\rm if} |\psi\rangle' = e^{if} 
|\psi\rangle \;,
\end{equation}
which induces a partition of the Hilbert space into equivalence classes, 
and then identifying points within each class to form a quotient space. For a Hilbert space ${\cal H}_n$ of complex dimension $n$, the associated projective space is the manifold $CP^{\,n-1}$, whose real dimension is $2n-1$. As the group of unitary transformations $U(n)$ acts transitively on ${\cal H}_n$, $CP^{\,n-1}$ can also be written as the quotient of $U(n)$ by the stability group associated with any state vector $|\psi_0\rangle $. Alternatively, noting that $\mathbb{U}= e^{i\phi}\,\bar{\mathbb{U}}$, with $\bar{\mathbb{U}} \in SU(n)$, $CP^{\,n-1}$ is the quotient of $SU(n)$ by the stability group 
$H \subset SU(n)$ that leaves the ket $|\psi_0\rangle $ invariant (up to a phase),
\begin{equation}
H = \{ h \in SU(n)~/~ h |\psi_0\rangle = e^{i\chi} |\psi_0\rangle \}\;.
\end{equation} 
As is well-known, $H$ is isomorphous to $U(n-1)$,
\begin{equation}
h = \left( \begin{array}{cc}
(\det u)^{-1}  & \mathbb{O}_{1 \times (n-1)} \\
\mathbb{O}_{(n-1) \times 1}  & u  \\
\end{array} \right) \makebox[.5in]{,} u \in U(n-1)
\;,
\end{equation} 
here, we have taken $|\psi_0\rangle$ as an $n\times 1$ column matrix 
whose first entry is the only nonvanishing element.  
Then, the space of physically distinct quantum states turns out to be, 
\begin{equation}
{CP}^{n-1}=SU(n)/U(n-1)\;.
\label{quot}
\end{equation}
This plays an important role in the kinematical approach of Mukunda and Simon, as the geometric phase is defined on the projective space, and its obtention  depends on computing the dynamical phase for a unitary evolution.

In a general topological space, closed paths can be separated into classes of homotopic paths that form the fundamental group $\Pi_1$. The group product is given by the class obtained from the composition of representative paths in each factor. In particular, closed paths in the projective space correspond to cyclic evolutions. However, the projective space $CP^{\,n-1}$ is not interesting from this point of view, as every closed path is topologically trivial,   
\begin{equation}
\Pi_1 (CP^{n-1}) = 0\;.
\label{trivial-topo}
\end{equation}
For example, for one qubit ($n=1$) the projective space $CP^1$ coincides with $S^2$, and every closed path on $S^2$ can be deformed to a point. A similar situation occurs for one qudit ($n=d$) and for a pair of qudits ($n=d^2$). As we will see, nontrivial 
topological properties in two-qudit systems may be manifested when the qudits can only undergo {\it local} unitary evolutions. Unlike the whole set of local and nonlocal unitary evolutions $U(d^2)$, local evolutions do not act transitively on the space of two-qudit states. Then, given an initial state, the evolutions can only explore a subspace characterized by a set of $d$ invariants, which lead to different subspaces and related topological properties. 

The most general two-qudit pure state $|\psi\rangle=\sum_{i,j=1}^{d} \alpha_{ij} | ij \rangle$ can be represented by the $d\times d$ matrix $\mathbf{\alpha}$ whose elements are the coefficients $\alpha_{ij}$. With this notation, the scalar product between two states is $\langle\phi|\psi\rangle=Tr(\mathbf{\beta^{\dagger}\alpha})$, where $\mathbf{\beta}$ represents $|\phi\rangle$. The total projective space 
$CP^{\,d^2-1}$ can be divided into subspaces labelled by a set of $d$ invariants,
\begin{equation}
I_p=Tr[(\alpha^\dagger \alpha)^{\, p}]=Tr[(\alpha\alpha^\dagger )^{\, p}] \;, 
\end{equation}
$p=1,\dots, d$, which due to the Cayley-Hamilton theorem determine the higher order correlators $I_p$, $p > d$. $I_1$ is simply the 
norm of the state vector, while $I_2$ is related to the {\it I-concurrence} of a two-qudit pure quantum state \cite{Iconc},
\begin{equation}
{\cal C}=\sqrt{2(1-I_2)}\;.
\end{equation}
The invariants $I_d$, $d\geq 2$, represent the well-known fact that entanglement is not affected by local unitary operations. In order to analyze the different subspaces, we note that any matrix $\alpha$ admits a singular value decomposition,
\begin{equation}
 \alpha = e^{i\phi} S_1 \Sigma S_2^T \;, 
\label{svd}
\end{equation}
where $S_k$, $k=1,2$ are in $SU(d)$ and $\Sigma$ is diagonal, with nonnegative real entries on the diagonal. Arranging $\Sigma|_{ii}=\sigma_i$ in descending order, $\Sigma$ is uniquely determined from $\alpha$. The quantitites $\sigma_i$, $i=1,\dots, d$ are  invariant under local unitary evolutions, they are related with the former invariants through,
\begin{equation}
I_p= \sum_i^d (\sigma^2_i)^{\, p} \;. 
\end{equation}
Unlike $\Sigma$, the $SU(d)$ matrices $S_1, S_2$ are not unique, so further analysis is needed to represent the different invariant projective subspaces 
$\pi(\sigma_1, \dots, \sigma_d)$, namely, the quotient spaces formed by equivalence classes of pairs of $SU(d)$ matrices, with the equivalence relation,
\begin{equation}
(S_1,S_2) \sim (S_1',S_2') \makebox[.5in]{\rm if}
S_1' \Sigma S_2'^T  = e^{if}\, S_1 \Sigma S_2^T\;.
\end{equation} 

\subsection{Separable states}
 
${\cal C}=0$ corresponds to separable states, as $Tr[\Sigma^2]=1$, $Tr[\Sigma^4]=1$ can only be satisfied by having one of the $\sigma_i$'s equal to $1$, and the rest vanishing. Then, in descending order we have, $\sigma_1=1$, $\sigma_i= 0$, for $i =2,\dots, d$, which gives,
\begin{equation} 
S_1 \Sigma S_2^T = u \otimes v
\makebox[.5in]{,} u= S_1 e_1 \makebox[.5in]{,} v = S_2 e_1 \;,
\end{equation}
where $e_i$ is a $d\times 1$ column matrix with elements $e_i|_j = \delta_{ij}$. As already discussed, $e_1$ is left invariant (up to a phase) by a subgroup of $SU(d)$ isomorphous to $U(d-1)$. In other words, the projective subspace 
$\pi (1,0,\dots,0)$ of zero concurrence states is simply $CP^{d-1} \otimes CP^{d-1}$, which is topologically trivial, 
\begin{equation}
\Pi_1 (CP^{d-1} \otimes CP^{d-1}) =\Pi_1 (CP^{d-1}) \oplus \Pi_1 (CP^{d-1}) =0\;.
\end{equation}

\subsection{Maximally entangled states}

On the other hand, let us consider the space of maximally entangled states, for which,
\begin{equation}
\Sigma =d^{-1/2}  I
\makebox[.5in]{,} 
I_p = 1/d^{p-1}  \;.
\end{equation}
In this case,
\begin{equation}
S_1 \Sigma S_2^T = d^{-1/2} S_m 
\makebox[.5in]{,} S_m = S_1 S_2^T \;,
\label{Sm}
\end{equation} 
so that the projective space is formed by the set of $SU(d)$ matrices $S_m$, with the identification,
\begin{equation}
S_m \sim S_m' \makebox[.5in]{\rm if}
S_m' = e^{if}\, S_m  \;.
\label{ids}
\end{equation}
As $S_m^{\prime}$ is also in $SU(d)$, the phase factor can only be in the center $Z(d)$, $f=2\pi z/d $, $z\in \mathbb{Z}$.
This identification is simply implemented by passing to the adjoint representation of $SU(d)$, which acts on the Lie algebra $\mathfrak{su}(d)$ according to,
\begin{equation}
X'= S_m X S_m^{-1} =  X'^A T_A
\makebox[.5in]{,}
X = X^A T_A \in \mathfrak{su}(d) \;.
\label{adj}
\end{equation}
Here, the $T_A$'s form a Lie algebra basis (for details, see section \ref{lie}).
This action can also be written in the form,
\begin{equation}
X'= X'^A T_A
\makebox[.5in]{,}
X'^A = R_{AB} X^B \makebox[.5in]{,} R=R(S_m) \;.
\end{equation}
From eq. (\ref{adj}), it is clear that $S_m$ and $ e^{2\pi i z/d}\, S_m $ represent one and the same adjoint transformation, $R(S_m)= R(e^{2\pi i z/d}\, S_m)$. This generalizes the well-known relation between $SU(2)$ and $Ad(SU(2))=SO(3)$, where
$S_m=I$ and $S_m=-I$ are both mapped into a trivial $SO(3)$ transformation $R=I_{3\times 3}$. Summarizing, the projective space of maximally entangled two-qudit states,
$$ \pi (d^{-1/2},d^{-1/2},\dots,d^{-1/2})\;,$$
is given by the adjoint representation of $SU(d)$,
\begin{equation}
Ad(SU(d)) = SU(d)/Z(d)\;,
\end{equation}
which is topologically nontrivial,
\begin{equation}
\Pi_1(Ad(SU(d))) = Z(d)\;. 
\end{equation}
Then, it becomes clear that depending on the values of the invariants, the topology of the projective subspaces can be modified.

\subsection{The  topology of rank $d$ states}

Here, it will be convenient changing to the polar decomposition,
\begin{equation}
\alpha = e^{i\phi}\, Q S_m \;,
\end{equation}
where $Q=S_1 \Sigma S_1^\dagger $ is hermitian and positive definite, while 
$S_m= S_1 S_2^T$. The projective subspace characterized by $I_p= Tr[(Q^2)^{p}]$ is then given by the set of matrices $Q S_m$, with the identification,
\begin{equation}
QS_m \sim Q'S_m' \makebox[.5in]{\rm if}
Q'S_m' = e^{if}\, QS_m \;.
\end{equation}
As $Q$, $Q'$ are hermitian, we must have $Q'=Q$, and the equivalence relation becomes,
\begin{equation}
QS_m \sim QS_m' \makebox[.5in]{\rm if}
QS_m' = e^{if}\, QS_m \;.
\label{einv}
\end{equation}
This is a general description of the projective subspaces, the solution to this relation is $S_m'=h S_m$, with $h\in H$, a subgroup of $SU(d)$ defined by,
\begin{equation}
Q h = e^{if}\, Q \;. 
\label{einv2} 
\end{equation}

Note that for separable states (rank $1$), $Q=u\otimes u^\ast$ is a $CP^{d-1}$ manifold, while the condition (\ref{einv2}) amounts to, 
\begin{equation}
u^\dagger h = e^{if}\, u^\dagger \;.
\end{equation}
Then, $H$ becomes isomorphous to $U(d-1)$, the equivalence classes of
$S_m$ matrices are labelled by points in $SU(d)/U(d-1)$, and
we make contact with the projective subspace  $CP^{d-1} \otimes\,  CP^{d-1}$. 
Other subspaces with rank lower than $d$ can be analyzed along similar lines.  
Here, we shall not discuss their classification. For rank $d$ states (nonzero $\sigma_i$'s) $Q$ is invertible, so $H=Z(d)$, and the projective space can be thought of as the set $Y$, formed by points
\begin{eqnarray}
y = Q S_m \;,
\end{eqnarray}
where (\ref{einv}) becomes the equivalence relation (\ref{ids}), the equivalence classes of $S_m$ matrices are labelled by points in $SU(d)/Z(d) = Ad(SU(d))$, and the invertible $Q$ can be written in terms of a hermitian traceless matrix $M$, 
\begin{equation}
Q = (\det Q)^{\frac{1}{d}}\, e^M \;.
\end{equation}
Note that there is a continuous map $F(y,s)$, defined on $Y\times [0,1] $,
\begin{equation}
F(y,s)= \frac{1}{[Tr\, Q^{\,2}(s)]^{\frac{1}{2}}}\, Q(s) \, S_m \makebox[.5in]{,} Q(s)= (\det Q)^{\frac{(1-s)}{d}} d^{-s/2}\,e^{(1-s)M}\;,
\end{equation}
with the following properties:
\begin{equation}
F(y,0)= Q\, S_m = y \makebox[.5in]{,} F(y,1)= d^{-1/2} \, S_m \;,
\end{equation}
while for points $a  \in A \subset Y$ representing maximally entangled states, that is $a=d^{-1/2} \, S_m$ ($Q=d^{-1/2} \, I$),
we have,
\begin{equation}
F(a,s)= d^{-1/2} \, S_m = a \makebox[.5in]{,} \forall \, s\in[0,1] \;.
\end{equation}
In addition, $y$ is identified with $y'$ if and only if 
$F(y,s)$ is identified with $F(y',s)$. These
properties mean that the projective space of maximally entangles states is a deformation retract of the projective space of rank $d$ states, so their first homotopy groups are equal \cite{24}, 
\begin{equation}
\Pi_1(Y) = \Pi_1 (A) = Z(d) \;.
\end{equation}

\section{Algebraic aspects of invariant subspaces}
\label{lie}
 
The Cartan decomposition of the Lie algebra $\mathfrak{su}(d)$ is an important tool that will be useful to characterize $\Sigma$, local evolutions, fractional phases, as well as monopole-like contributions to the geometric phase.
Let us consider an $\mathfrak{su}(d)$ basis $T_{A}$, $A=1,2,...,d^2-1$, satisfying, 
\begin{equation}
[T_A,T_B]=if_{ABC} T_{C} \makebox[.5in]{,}
Tr(T_A T_B)= \frac{1}{2d}\, \delta_{AB} \;,
\label{Lie-alg}
\end{equation}
where $f_{ABC}$ are the structure constants. They form
a basis for the space of Hermitian $d\times d$ traceless matrices, and can be separated into diagonal elements $T_q$, $q=1,\dots, d-1$, which generate a Cartan subgroup $H=U(1)^{d-1}$, and $d(d-1)$ off-diagonal elements. The latter can be written as,
\begin{equation}
T_{\alpha}=\frac{1}{\sqrt{2}}(E_\alpha + E_{-\alpha})
\makebox[.5in]{,}
T_{\bar{\alpha}}=\frac{1}{\sqrt{2}i}(E_\alpha - E_{-\alpha})  \;,
\label{Tes}
\end{equation}
where the nonhermitian $E_{\alpha}, E_{-\alpha}$ satisfy,
\begin{equation}
[T_q,T_p]=0 \makebox[.3in]{,}
[T_q,E_{\alpha}]=\vec{\alpha}|_q\, E_{\alpha}
\makebox[.3in]{,}
[E_{\alpha},E_{-\alpha}]= \vec{\alpha}|_q\, T_q\;.
\label{algebrag}
\end{equation}
The subindex $\alpha$ indicates a positive $(d-1)$-tuple $\vec{\alpha}$
(positive root) whose $\vec{\alpha}|_q$ component is defined by the previous commutators\footnote{An $r$-tuple is defined as positive if the last nonvanishing component is positive.}.
Now, as $\Sigma^2$ is diagonal, it can be written in the form,
\begin{equation}
\Sigma^2 = d^{-1} I + b_q T_q\;,
\end{equation}
where we already used the normalization condition $I_1=Tr [\Sigma^2] =1$..
Note that \cite{12}
\begin{equation}
{\cal C}=\sqrt{2(1-Tr[\Sigma^4])} \makebox[.5in]{,} 
 Tr[\Sigma^4] = \frac{1}{d} + \frac{b^2}{2d} 
\makebox[.5in]{,}
b= \sqrt{b_q b_q}
\;,
\end{equation}
so that the maximum value of the concurrence occurs for $b^2=0$, thus giving
\begin{equation}
b = \sqrt{d({\cal C}_m^2-{\cal C}^2)} 
\makebox[.5in]{,}
{\cal C}_m=\sqrt{2(d-1)/d}   \;.
\label{Q2}
\end{equation}
The coefficients $b_q$ have a nice interpretation in terms of weights.  A weight $\vec{w}$ is defined by the eigenvalues of diagonal generators corresponding to one common eigenvector. 
In particular, the second equation in (\ref{algebrag}) says that the roots are the weights of the adjoint representation, which acts via commutators. 
In the fundamental representation, the diagonal of $T_q$ can be given by,
\begin{equation}
\frac{1}{\sqrt{2q(q+1)d}}\,(1,\dots,1,-q,0,\dots,0)\;,
\end{equation}
where the initial $q$ elements are equal to $1$.  The weights of the fundamental representation, $\vec{w}_{i}$, $i=1,\dots,d$, are then given by,
\begin{equation}  
\vec{w}_{i} = (T_1|_{ii}, T_2|_{ii},\dots, T_{d-1}|_{ii}) \;,
\label{wfun}
\end{equation}
and satisfy \cite{28}, 
\begin{equation}
\vec{w}_1+\dots +\vec{w}_d=0\makebox[.3in]{,}
\vec{w}_i\cdot \vec{w}_i = \frac{d-1}{2d^2}
\makebox[.3in]{,}
\vec{w}_i \cdot \vec{w}_j = -\frac{1}{2d^2}
\makebox[.3in]{,}
i\neq j\;.
\label{fundwei}
\end{equation}
Then, it becomes clear that,
\begin{equation}
b_q = 2d\, Tr (T_q \Sigma^2) = 2d\, \sum_{i=1}^d \sigma_i^2\, T_q|_{ii} = \sum_{i=1}^d \sigma_i^2\, 
\vec{\beta}_i|_{q}   \;,
\end{equation}
where the ``magnetic weights'' of the fundamental representation of $\mathfrak{su}(d)$ are defined by, 
\begin{equation}
\vec{\beta}_i = 2d \vec{w}_i \;.
\end{equation}

\subsection{Local evolutions and non Abelian connections}

Under local unitary evolutions, we have,
\begin{equation}
 \alpha (t)= e^{i\phi(t)} S_1(t) \Sigma S_2^T(t)=e^{i\phi(t)} Q(t) S_m(t) \;,
 \label{locev}
\end{equation}
where $\Sigma$ is time-independent. The evolution equation can be written as,
\begin{equation}
-i\dot{\alpha} = \dot{\phi}\, \alpha  + A_1\, \alpha + \alpha\, A_2^T \;,
\end{equation}
where $A_1$, $A_2$ are defined by,
\begin{equation}
A= i S \frac{d~}{dt} S^{-1}
\;,
\label{stronger}
\end{equation}
for $S=S_1, S_2$, respectively. It will be convenient introducing a 
time-dependent  Lie basis,
\begin{equation}
n_A= S T_A S^{-1} \makebox[.5in]{,} [n_A,n_B]=if_{ABC}\, n_C
\;,
\label{vman-0}
\end{equation}
and represent the single-qudit ``hamiltonian'' $A$, given in eq. (\ref{stronger}), using these variables. The evolution $n_A(t)$ can be thought of as occuring in the adjoint representation of $SU(d)$. In components with respect to the basis $T_A$, we can write,
\begin{equation}
n_A=\hat{n}_A\cdot \vec{T} \,,~~ \hat{n}_A=R(S)\, \hat{e}_A\,,~~ R \in Ad(SU(d))\;,
\label{vman}
\end{equation}
where $\hat{e}_A$ is a $(d^2-1)\times 1$ matrix, with elements $\hat{e}_A|_B=\delta_{AB}$. Now, defining the covariant derivative,
\begin{equation}
D \psi= \dot{\psi} -i[A,\psi]\;,
\label{cov-abs}
\end{equation}
where $A$ and $\psi$ are in the Lie algebra, it is easy to see that $n_A$ satisfies,
\begin{equation}
D n_A = 0 \;,
\end{equation}
which implies 
\begin{equation}
[n_A, Dn_A]=0\;,
\end{equation}
where the repeated index is summed over $A=1,\dots,d^2-1$. 
Now we recall that, in the algebra, a positive definite metric exists,
\begin{equation}
\langle X ,Y\rangle =Tr \left(Ad(X) Ad(Y)\right)\;,
\end{equation}
where $Ad(X)$ is a linear map of $X\in \mathfrak{su}(d)$ into the adjoint representation generated by the $d\times d$ hermitian matrices 
 $M_A$, with elements $M_A|_{BC}=-i f_{ABC}$, satisfying,
\begin{equation}
\left[M_{A},M_{B}\right]=if_{ABC} M_C\;,
\label{algeb}
\makebox[.5in]{,}
Tr(M_A M_B)=\delta_{AB}\;.
\label{alg-norm}
\end{equation}
With this normalization, we have, 
\begin{equation}
f_{ABC}\, f_{DBC} =\delta_{AD} 
\makebox[.5in]{,} \langle T_A,T_B\rangle = \delta_{AB}
\makebox[.5in]{,}
\langle n_A , n_B \rangle =\delta_{AB} \;.
\label{norma}
\end{equation} 
Then, using the 
property,
\begin{equation}
n_A \wedge (X \wedge n_A) 
= \langle X , n_B\rangle\, n_A \wedge (n_B \wedge n_A) = X\;,
\end{equation}
where\footnote{As we are using hermitian generators, we define the closed product in the Lie algebra,
\begin{equation}
X \wedge Y = -i [X,Y]\;.
\end{equation}},
\begin{equation}
n_A \wedge n_B = f_{ABC}\, n_C \;,
\end{equation}
we get,
\begin{equation}
i S \frac{d~}{dt} S^{-1} = -n_A \wedge  \dot{n}_A \;.
\label{rep1}
\end{equation}
Finally, we can expand $\dot{n}_A= \langle n_B,\dot{n}_A \rangle \, n_B$ to write,
\begin{eqnarray}
i S \frac{d~}{dt} S^{-1} = -C_A\, n_A 
\makebox[.5in]{,}
C_A = -f_{ABC}\, \langle n_B,\dot{n}_C \rangle \;,
\label{rep2}
\end{eqnarray}
(see also ref. \cite{23}).

\subsection{Fractional phases}
\label{f-ph} 

We shall consider closed paths in the projective space given by, 
\begin{equation}
S(0)=I\makebox[.5in]{,} S(\tau)=e^{2\pi i z/d}\, I  \makebox[.5in]{,} 
z \in \mathbb{Z} \;,
\end{equation}
In order to implement these conditions, we introduce the coset decomposition,
\begin{equation}
S = U V \makebox[.5in]{,}
U\in SU(d)/U(1)^{d-1}
\makebox[.5in]{,}
V\in U(1)^{d-1}\;, 
\label{uv}
\end{equation}
where $U(1)^{d-1}$ is the subgroup of $SU(d)$ generated by $T_q$, $q=1, \dots, d-1$.
The coset $U$-sector can be defined by the following requirement: if for every diagonal generator,
\begin{equation}
U T_q U^{-1} = T_q\;,
\label{req}
\end{equation}
then necessarily $U=I$.  Then, the fractional phase must be generated in the $V$-sector.  This is because $e^{2\pi i z/d}\, I $ satisfies eq. (\ref{req}) but it is nontrivial.
Then, the cyclic evolution is given by, 
\begin{equation}
U(0)=I\makebox[.5in]{,} U(\tau)= I  \;,
\label{ccon} 
\end{equation}
\begin{equation}
V(t)= e^{i\, h_q(t) T_q}  \makebox[.5in]{,} h_q(0)=0 \makebox[.5in]{,} h_q(\tau) = 2\pi\vec{\beta}|_q \;,
\label{vv}
\end{equation}
\begin{equation}
e^{i2\pi\, \vec{\beta} \cdot \vec{T}} = \mathfrak{z}\, I_{d\times d} 
\makebox[.5in]{,}
\mathfrak{z}=e^{2\pi i z/d}
\;.
\label{poss-b} 
\end{equation}
Here, we adhered to the convention that in dot products involving a weight, $\vec{\beta}$ is considered as $(\beta_1,\dots, \beta_{d-1},0,\dots,0)$.
The possible $\vec{\beta}$'s have been discussed in refs. 
\cite{18}-\cite{22}, following the ideas introduced in \cite{25} to characterize non Abelian monopoles.
They are given by,
\begin{equation}
\vec{\beta}=2d\vec{w}\;,
\label{betasw} 
\end{equation}
where $\vec{w}$ are the weights of any representation of $SU(d)$. 
The fundamental and antifundamental weights correspond to the minimum charges \cite{18}. In this respect, note that,
\begin{equation}
 e^ {i2\pi\, \vec{\beta} \cdot \vec{T}}\, e_j = e^ {i2\pi\, \vec{\beta} \cdot
 \vec{w}_j}\, e_j\;.
 \label{so}
\end{equation}
When $\vec{\beta}$ is a fundamental magnetic weight $\vec{\beta}_i$, as $2\pi\, \vec{\beta}_i \cdot \vec{w}_j$ corresponds to either $- \frac{2\pi}{d}$, if $i\neq j$, or $\frac{2\pi(d-1)}{d}\equiv - \frac{2\pi}{d}$, if $i=j$ (cf. eq. (\ref{fundwei})), a value $z=-1$ is implied. The generators of the antifundamental representation $-T_A^\ast$ have associated weights $-\vec{\beta}_i$, thus leading to $z=1$. 
 
\section{Algebraic aspects of the geometric phase} 

The geometric phase for a general nonadiabatic and noncyclic evolution is given by \cite{10,11}, 
\begin{eqnarray}
\phi_g & =& \arg{\langle\psi(0)|\psi(\tau)\rangle} + 
i\int_0^\tau dt \,\,\langle\psi(t)|\dot{\psi}(t)\rangle \;.
\label{phig}
\end{eqnarray}
This phase is gauge invariant under the state vector transformation,
\begin{equation}
|\psi(t)\rangle \to e^{if(t)}|\psi(t)\rangle \;,
\end{equation}
and is also invariant under reparametrizations, only depending on the path the system follows on the projective space. Now,  for a qudit pair undergoing local unitary evolutions, the total projective space can be replaced by the invariant subspace $\pi(\sigma_1, \dots, \sigma_d)$.
Because of gauge invariance, the continuous phase $\phi(t)$ in eq. (\ref{locev}) does not contribute to $\phi_g$. It is important to underline that, depending on the rank of the state, other phases $e^{if(t)}$ could be generated {\it all along} the evolution. Take for example an evolution operated on the first qudit 
($S_2(t) \equiv I$),  
\begin{equation}
 \alpha (t)= e^{i\phi(t)} S_1(t) \Sigma  \;,
\end{equation}
where $S_1=U_1 V_1$, with $U_1,V_1$ satisfying eqs. (\ref{ccon})-(\ref{poss-b}). 
For a separable state, $\Sigma = e_1 \otimes e_1$, 
\begin{equation}
 \alpha_{\rm rank-1} (t)= e^{i\phi(t)}e^{if(t)}  U_1(t) \Sigma \makebox[.5in]{,} 
e^{if(t)} = e^{i\, h_q(t) T_q|_{11}} \;.
\end{equation}
Then, because of gauge invariance, although the $V$-sector generates a fractional phase at time $\tau$, this sector does not contribute to the geometric phase for separable states. On the other hand, for rank $d$ states,
\begin{equation}
S_1(t) \Sigma = e^{if(t)} \Sigma \makebox[.5in]{\rm iff} S_1(t) = e^{if(t)} I\;,
\end{equation}
so that $e^{if(t)}$ must be a center element. As the possibilities are discrete, such phase factor can only be attained at the end of the cyclic evolution ($t=\tau$), so the $V$-sector is expected to contribute to $\phi_g$ in this case. This can be explicitly verified by computing $\phi_g$, using the general algebraic properties discussed above. From eq. (\ref{phig}), it is easy to see that,
\begin{equation}
\phi_g = Tr (\alpha^\dagger(0) \alpha (\tau)) + \phi(S_1) + \phi(S_2) \;,
\label{e1}
\end{equation}
\begin{eqnarray}
\phi(S)&=&  i\int_0^\tau dt \, Tr  \left(\Sigma^2 S^{-1} \frac{dS}{dt}\right) \nonumber \\
&=& \sum_{i=1}^d \sigma_i^2\, 
\vec{\beta}_i|_q \,  \Phi_q(S) \makebox[.5in]{,} \Phi_q(S)=i\int_0^\tau dt \, Tr\left[T_q\,S^{-1} \frac{dS}{dt}\right]\;,
\label{geomphase}
\end{eqnarray}
where $S(t)$ represents anyone of the evolutions $S_1(t)$ and $S_2(t)$ in eq. 
(\ref{locev}). In addition, we can use the decomposition in eqs. (\ref{uv}), (\ref{vv}) to obtain,
\begin{equation}
\Phi_q(S)=\Phi_q(V) + \Phi_q(U) \makebox[.5in]{,}
\Phi_q(V)= -\frac{ h_q(\tau)}{2d} \;.
\label{e2}
\end{equation}
Then, for cyclic evolutions,
\begin{equation}
\phi_g = \phi_g(S_1) + \phi_g(S_2)  \;,
\end{equation}   
\begin{equation}
\phi_g(S) =\frac{2\pi z }{d} -2\pi  \sum_{j=1}^d \sigma_j^2\, 
  \vec{w}_j \cdot \vec{\beta} + \sum_{j=1}^d \sigma_j^2\, 
\vec{\beta}_j|_q\Phi_q(U)\;.
\label{s4}
\end{equation}
For maximally entangles states, 
$\sigma_1= \dots =\sigma_d = 1/d^{-1/2} $, so that using the first equation in (\ref{fundwei}), we get, 
\begin{equation}
\phi_g(S) =\frac{2\pi z }{d} \;.
\end{equation}
In general, recalling that for a closed path characterized by a fractional phase $e^{\frac{2\pi i z}{d}}$, $z=\mp 1$, $\vec{\beta}$ is a fundamental (antifundamental) weight 
$\vec{\beta}_i$  ($-\vec{\beta}_i$) (see the discussion after eq. (\ref{so})), we get,
\begin{equation}
\phi_g(S) = \mp 2\pi \sigma_{i}^2 + \sum_{j=1}^d \sigma_j^2\, 
\vec{\beta}_j|_q\Phi_q(U)\;,
\label{gint}
\end{equation}
where we used the normalization condition $\sum_{j=1}^d \sigma_j^2 =1$.
Note that for a separable state $\sigma_1=1$, and $\sigma_i =0$, for $i\neq 1$, so that the contribution originated from the $V$ sector is an unobservable $0$ or $\pm 2\pi$ phase, as anticipated. 

\subsection{The monopole-like coset contribution}

Our next objective is representing the coset contribution to the geometric phase in a form that generalizes the one for a single qubit. In the latter case, the projective space is $S^2$ (Bloch sphere) and, for cyclic evolutions, $\phi_g$ is given by half the solid angle subtended by the path from the origin, a purely geometrical effect. Let us recall the main steps of the single qubit calculation. Considering an initial state $e_1$, 
the evolution is,
\begin{equation}
|\psi(t) \rangle  = e^{i\phi(t)}\,S(t) 
\left( \begin{array}{cc}
1  \\
0  \\ 
\end{array}\right)
\makebox[.5in]{,}
S(t)=\left( \begin{array}{cc}
\alpha(t) & \gamma(t) \\
-\bar{\gamma}(t) & \bar{\alpha}(t) \\
\end{array} \right) 
\makebox[.5in]{,} \alpha \bar{\alpha} + \gamma \bar{\gamma}=1 \;.
\end{equation}
By writing $\alpha= |\alpha|\, e^{i\chi}$, $\gamma = \beta \, e^{-i\chi}$, we have $S=UV$,
\begin{equation}
U=\left( \begin{array}{cc}
|\alpha| & \beta \\
-\bar{\beta} & |\alpha| \\
\end{array} \right) 
\makebox[.5in]{,}
V=\left( \begin{array}{cc}
e^{i\chi} & 0 \\
0 & e^{-i\chi} \\
\end{array} \right) \;.
\end{equation}
Then, $|\psi(t) \rangle  = e^{i(\phi(t)+\chi(t))}\, U(t) e_1$ and, due to gauge invariance, the geometric phase is a function defined on $U$, a point in $SU(2)/U(1)$.
Using the parametrization, 
\begin{equation}
U=\left( \begin{array}{cc}
\cos (\theta/2) & +\sin(\theta/2)\, e^{i\varphi} \\
-\sin(\theta/2)\, e^{-i\varphi}  & \cos (\theta/2) \\
\end{array} \right) 
\;,
\label{Upar}
\end{equation}
$0\leq \theta \leq \pi$, $0\leq \varphi \leq 2\pi$, together with Green's theorem, the geometric phase
\begin{equation}
\phi_g = \frac{1}{2} \int d\varphi d\theta \sin \theta = \frac{\Omega}{2} \;,
\end{equation}
is obtained, which is half a solid angle over an $S^2$-sphere. At this stage, looking at eq. (\ref{Upar}), it is still not clear that the projective space for a single qudit is $S^2$, as the quantities $\cos (\theta/2)$, $\sin(\theta/2)\, e^{i\varphi}$ seem to parametrize half a sphere. However, at $\theta =\pi$, the kets 
$U(\varphi) \,e_1 $ are physically equivalent for every $\varphi$. The identification of all these points is what leads to the $S^2$ manifold. A nice manner to implement this identification is by the map $U\sigma_3 U^{-1}=\hat{n}\cdot \vec{\sigma}$.  Different points on the manifold $\hat{n}\in S^2$ describe physically distinct states. In these variables, the Berry phase can be represented as,
\begin{equation}
\phi_g = \frac{1}{2} \int d\theta \, d\varphi\,  \hat{n} \cdot 
\left(\frac{\partial \hat{n}}{\partial \theta} \times \frac{\partial \hat{n}}{\partial  \varphi} \right) \;,
\label{unmedio}
\end{equation}
see \cite{26}. Now, we would like to generalize this type of representation to
the coset contribution of a general two-qudit system. In this case, the natural variables are expected to be the local Cartan basis elements,
\begin{equation}
n_q = U T_q U^{-1} = S T_q S^{-1} \;.
\end{equation}
In general, the $U$ sector can be written in terms of the factorization \cite{27,Ak},
\begin{equation}
U = \Omega^{(d;d)}\, \Omega^{(d-1;d)} \dots\, \Omega^{(2;d)}\;, 
\label{sectors}
\end{equation}
where the coset representatives have the structure,
\begin{equation}
\Omega^{(m;d)}=\left( \begin{array}{cc}
SU(m)/U(m-1) & \mathbb{O}_{m \times (d-m)} \\
\mathbb{O}_{(d-m) \times m}  & \mathbb{I}_{(d-m)\times (d-m)} \\
\label{sst}
\end{array} \right) 
\;.
\end{equation} 
The time evolution can be given in terms of a set of parameters, in the form 
$U=U(P_\mu)$, $P_\mu=P_\mu(t)$,
$i=1,\dots, {\cal D}$,  which because of (\ref{ccon}) describe a closed loop. Therefore, 
\begin{equation}
\Phi_q(U)= \frac{1}{2d} \int_0^T dt \, {\cal C}_q = \frac{1}{2d} \oint_\Gamma dP_\mu \, {\cal C}^q_\mu \;.
\end{equation}
Similarly to eq. (\ref{rep2}), we can write,
\begin{equation}
i U \partial_\mu U^{-1} = -{\cal C}^A_\mu\, u_A 
\makebox[.5in]{,}
{\cal C}^A_\mu=  -f_{ABC}\, \langle u_B,\partial_\mu u_C \rangle \makebox[.5in]{,} u_A=  U T_A U^{-1} \;,
\label{ueq}
\end{equation}
where $\partial_\mu$ is the partial derivative with respect to $P_\mu$. Then, using Stoke's theorem, we get,
\begin{eqnarray}
\Phi_q(U)&=& \frac{1}{2d} \int_{S(\Gamma)} dS_{\mu \nu}\, (\partial_\mu {\cal C}^q_\nu -\partial_\nu {\cal C}^q_\mu)
\nonumber \\
&=& \frac{1}{2d} \int_{S(\Gamma)} dS_{\mu \nu}\, (F_{\mu \nu}^q({\cal C}) - H^q_{\mu \nu}({\cal C}))\;,
\label{phiq}
\end{eqnarray}
where $F^A_{\mu \nu}$ is the usual non Abelian field strength tensor,
\begin{equation}
F^A_{\mu \nu}({\cal C})=\partial_\mu {\cal C}^A_\nu-\partial_\nu {\cal C}^A_\mu +  f^{ABC} {\cal C}^B_\mu {\cal C}^C_\nu\;,
\end{equation}
and we have defined,
\begin{equation}
H^q_{\mu \nu}({\cal C})= f_{qBC} {\cal C}^B_\mu {\cal C}^C_\nu \;.
\label{eum}
\end{equation} 

For a general connection $A^A_\mu T_A$, the field strength tensor is given in terms of the commutator of covariant derivatives. As 
\begin{equation}
F^A_{\mu\nu}(A) T_A =i\left[D_\mu,D_\nu \right]
\makebox[.3in]{,}
D_\mu=\partial_\mu-i A^A_\mu T_A\;,
\end{equation}
taking into account eq. (\ref{ueq}), or equivalently,
\begin{equation}
 iU^{-1} \partial_\mu U  = {\cal C}^A_\mu \, T_A\;.
\label{Sder}
\end{equation} we have ($U(P_\mu)$ is single-valued),
\begin{equation}
F^A_{\mu\nu}({\cal C})= i\, {\rm tr}\, (T^A
U^{-1}[\partial_\mu,\partial_\nu]U) =0 
\;,
\label{FCe}
\end{equation}
\begin{equation}
\Phi_q(U) = -\frac{1}{2d} \int_{S(\Gamma)} dS_{\mu\nu}\,  H^q_{\mu\nu}({\cal C})\;.
\label{o4}
\end{equation}

With these tools, we are ready to obtain an expression for the coset contribution.
The mathematics involved is the same appearing when 
analyzing junctions formed by center vortices and monopoles in Yang-Mills-Higgs models (see ref. \cite{23}, and references therein). Initially, we note that,
\begin{eqnarray}
\langle u_B , \partial_\mu u_C\rangle  &=& \hat{u}_B\cdot \partial_\mu \hat{n}_C =  \hat{e}_B \cdot ({\cal R}^{-1}\partial_\mu {\cal R}\, \hat{e}_C ) \nonumber \\
&=& {\cal R}^{-1}\partial_\mu {\cal R}|_{BC} \;.
\end{eqnarray}
Using the adjoint version of  eq. (\ref{Sder}),
\begin{eqnarray}
i  {\cal R}^{-1} \partial_\mu {\cal R} = {\cal C}^A_\mu\, M_A
\makebox[.5in]{,} {\cal R} = R(U)\;,
\end{eqnarray}
we have, 
\begin{equation}
 \langle u_B , \partial_\mu u_C\rangle= -i{\cal C}^A_\mu \, M_A|_{BC} =- f_{ABC}\, {\cal C}_\mu^A  \;,
\label{Cprop}
\end{equation}
thus obtaining, 
\begin{eqnarray}
\langle u_C , \partial_\mu u_A \wedge \partial_\nu u_B\rangle &=&
\langle u_C , u_{A'}\wedge u_{B'} \rangle \langle u_{A'}, \partial_\mu u_A\rangle \langle u_{B'}, \partial_\nu u_B\rangle \nonumber \\
&=& i M_A M_C M_B|_{DE} \,{\cal C}^D_\mu  {\cal C}^E_\nu \;,
\end{eqnarray}
\begin{equation}
\langle  P_{\beta}, \partial_\mu P_{\beta} \wedge \partial_\nu P_{\beta} \rangle
= i   (\vec{\beta} \cdot  \vec{M})^3|_{DE} \, {\cal C}^D_\mu  {\cal C}^E_\nu
\makebox[.5in]{,}
P_\beta=\vec{\beta}|^q u_q
 \;.
\end{equation}
Next, for a magnetic weight $\vec{\beta}_j = 2d\vec{w}_j$, where $\vec{w}_j$ is a fundamental weight,
\begin{equation}
(\vec{\beta}_j\cdot  \vec{M})^3 =  \vec{\beta}_j\cdot  \vec{M}\;.
\label{qu-lin}
\end{equation}
This can be seen as follows. The second equation in (\ref{algebrag}) tell us that the roots are weights of the adjoint representation. Indeed, this statment can be obtained in matrix notation, by expanding 
$E_\alpha ={\cal E}_\alpha^A T_A$, using the Lie algebra (\ref{Lie-alg}), and the  fact that $M_q|_{AB} = -i f_{qAB}$,
\begin{equation}
M_q \, {\cal E}_\alpha =\vec{\alpha}|_q\, {\cal E}_\alpha\;, 
\end{equation}
where ${\cal E}_\alpha$ is a $(d^2-1)\times 1$ column matrix with components 
${\cal E}_\alpha^A$. That is,
$$(\vec{\beta}_j\cdot  \vec{M})^3 \, {\cal E}_{\pm \alpha} = \pm (\vec{\beta}_j \cdot  \vec{\alpha})^3 \, {\cal E}_{\pm \alpha}  \;.$$ 
In addition, the nonzero roots can be written as 
$\vec{w}_i -\vec{w}_k$, $i \neq k $ (see ref. \cite{28}), which together with the properties
(\ref{fundwei}) imply that the only possible values for $\vec{\beta}_j \cdot  \vec{\alpha}$ are $-1,0,+1$. Namely,
$$(\vec{\beta}_j\cdot  \vec{M})^3 \, {\cal E}_{\pm \alpha} = (\vec{\beta}_j\cdot  \vec{M}) \, {\cal E}_{\pm \alpha}  \;.$$ 
Similarly, the first equation in (\ref{algebrag}) gives,
$$(\vec{\beta}_j\cdot  \vec{M})^3 \, {\cal T}_q = (\vec{\beta}_j\cdot  \vec{M}) \, {\cal T}_q= 0 \;,$$
where ${\cal T}_q$ is a $(d^2-1)\times 1$ column matrix with the $q$-th element
equal to $1$ and other elements equal to zero. In this manner, we have shown eq.
(\ref{qu-lin}), as both members give the same result when applied on a basis. 
This,  together with eq. (\ref{eum}), leads to, 
\begin{equation}
\langle P_{\beta}, \partial_\mu P_{\beta} \wedge \partial_\nu P_{\beta} \rangle= \vec{\beta}|_q f_{qDE} \,{\cal C}^E_\mu  {\cal C}^E_\nu =  -\vec{\beta}|_q H^q_{\mu\nu}({\cal C})  
 \;.
\label{topo-curr}
\end{equation}
Finally, using  eq. (\ref{o4}), the coset contribution in eqs. (\ref{s4}), (\ref{gint}) can be cast in the form,
\begin{eqnarray}
\sum_{j=1}^d \sigma_j^2\, 
\vec{\beta}_j|_q\Phi_q(U) &=& -\frac{1}{2d}\sum_{j=1}^d \sigma_j^2 \int_{S(\Gamma)} dS_{\mu\nu}\,
\vec{\beta}_j|_q    H^q_{\mu\nu}({\cal C}) \nonumber \\
& = & \frac{1}{2d} \sum_{j=1}^d \sigma_j^2\int_{S(\Gamma)} dS_{\mu\nu}\, \langle P_{\beta_j}, \partial_\mu P_{\beta_j} \wedge \partial_\nu P_{\beta_j} \rangle\;.
\label{solidang}
\end{eqnarray}  
This is a sum, weighted by the $\sigma^2_j$ invariants, of quantities that generalize the well-known expression for a single qubit (cf. eq. (\ref{unmedio})). This formula can be checked for entangled qubits ($d=2$), where the weights are one component. Those of the fundamental representation are $\omega_1=\frac{1}{2\sqrt{2}}$, $\omega_2=-\frac{1}{2\sqrt{2}}$, so that the coset contribution (\ref{solidang}) becomes,
\begin{equation}
(\sigma_1^2-\sigma_2^2)\, \frac{\Omega}{2} \;,
\label{accor} 
\end{equation}
where $\Omega$ is the solid angle subtended by the path $\hat{n}(t)\in S^2$, $S\sigma_3 S^{-1}=\hat{n}\cdot \vec{\sigma}$, from the origin. The formula in eq. (\ref{solidang}) can also be checked for an evolution that is restricted to an $SU(2)$ subgroup of $SU(d)$. In this respect, the following triplets,
\begin{equation}
\frac{1}{\alpha^2}\, \vec{\alpha}\cdot
\vec{T}\makebox[.5in]{,}
\frac{1}{\sqrt{\alpha^2}}\, T_{\alpha}
\makebox[.5in]{,}
\frac{1}{\sqrt{\alpha^2}}\, T_{\bar{\alpha}}\;,
\label{subalg}
\end{equation}
generate $\mathfrak{su}(2)$ subalgebras
of $\mathfrak{su}(d)$ (cf. eqs. (\ref{Tes}), (\ref{algebrag})). With the restriction, the coset reduces to $S^2$. Then, similarly to the qubit case, the integral $\Phi_q(U)$ can be thought of as a magnetic flux computed through a surface contained on $S^2$, as if we had a magnetic monopole placed at the center of $S^2$. The total flux of such configuration is $2d\, 2\pi\, \vec{\alpha}|_q$ 
(see ref. \cite{23}), so for restricted evolutions, 
\begin{equation}
\Phi_q(U)= 2d\, \frac{\Omega}{2}\, \vec{\alpha}|_q\;.
\end{equation}
Then, using again that the roots $\vec{\alpha}$ can be written as weight differences,
together with eq. (\ref{fundwei}), for $\vec{\alpha}= \vec{\alpha}_{ik} = \vec{w}_i -\vec{w}_k$ ($i \neq k $), the coset $U$-contribution to the 
geometric phase becomes,
\begin{eqnarray}
\sum_{j=1}^d \sigma_j^2\, 
\vec{\beta}_j|_q\Phi_q(U) &=& \frac{\Omega}{2}\, \sum_{j=1}^d \sigma_j^2\, 
\vec{\beta}_j\cdot \vec{\alpha} = (\sigma_i^2 - \sigma_k^2)\, \frac{\Omega}{2}  \;,
\end{eqnarray}
in consonance with eq. (\ref{accor}). For general $SU(d)$ evolutions, all the sectors
$CP^{m-1}= SU(m)/U(m-1)$, $2\leq m \leq d$, in eqs. (\ref{sectors}), (\ref{sst}) will contribute to the geometric phase. A ``solid angle'' formula for the Berry phase in a three-level system, which is defined on $SU(3)/U(2)$, has been developed in refs. \cite{simon, byrd}. The monopole-like contributions in eq. (\ref{solidang}) could be helpful to understand generalized solid angles in terms of the paths followed 
by $n_q$. These are natural variables that uniquely determine a point on the different projective spaces.

\section{Conclusions}

The  generation of geometric and topological phases in quantum systems constitute a fascinating area of research. The possibility of having a physical output that is insensitive to details of the evolutions, only depending on trajectories followed in parameter space, or classes of homotopic trajectories, is attractive to implement quantum gates. The cleaner the output, the better the operations could be protected against noise and decoherence.

In this work, we have further developed the understanding of these phases in the context of entangled $d$-level states. Initially, we analyzed the topology of some projective subspaces that are invariant under unitary local evolutions. While the first homotopy group of the projective space for separable states is trivial, that for maximally entangled states (MES) is $Z(d)$. 

The projective space for general rank-$d$ states also has a topology characterized by $Z(d)$, as it can be deformed by a retraction to the MES projective space. This space is obtained by an identification of state vectors differing by a fractional phase. Topologically nontrivial closed paths are precisely those generating the fractional phases. For MES states, this means that the projective space is given by the adjoint representation of $SU(d)$.  Of course, a fractional phase can also be generated for separable states. The  difference is that it is built as a continuous phase factor $e^{if(t)}$ all along the evolution and, because of gauge invariance, this part of the evolution cannot contribute to the geometric phase.

The general structure of the $\mathfrak{su}(d)$ Lie algebra provides the appropriate mathematical tools  to characterize two-qudit states, fractional and geometric phases. Local evolutions can be separated into coset and Cartan sectors. In the Cartan sector, the fractional phases are attained at some points in parameter space that form the lattice of weights of the different $SU(d)$ representations. The simplest fractions, $e^{\pm 2\pi i/d}$, correspond to the weights of the fundamental and antifundametal representations. These properties also play an important role in another physical context, namely, when characterizing smooth center vortex solutions in Yang-Mills-Higgs models. 

As is well-known, the projective space for a two-level system is $CP^1=SU(2)/U(1)$, which corresponds to an $S^2$ manifold. This manifold is obtained from the map 
$S \sigma_3  S^{-1}=\hat{n}\cdot \vec{\sigma}$, $\hat{n}\in S^2$, where $\sigma_3$ is the diagonal Pauli matrix. In that case, the geometric phase is given by half the  solid angle subtended by $\hat{n}$ from the origin. This in turn can be written as
the flux of a topological charge density on $S^2$, for an $S^2 \to \hat{n}\in  S^2$ mapping, a monopole-like contribution. In our work, by using a time-dependent Lie algebra basis for $\mathfrak{su}(d)$, we carefully showed how to identify monopole-like contributions for qudits. This was done by introducing local Cartan elements $n_q$, $q=1,\dots, d-1$, projected along the fundamental weights of $\mathfrak{su}(d)$. For entangled qudits, the coset part of the geometric phase was finally obtained as a sum of the partial contributions weighted by $d$ invariants under local unitary evolutions.

Further studies could be oriented to understand all possible topologies as a function of the invariants, ranging from rank-$1$ to rank-$d$ states. These ideas could shed light on intrinsically different behaviours of quantum systems as a function of parameters and their corresponding topologies. 

\section*{Acknowledgments}

Funding was provided by 
Conselho Nacional de Desenvolvimento Tecnol\'ogico (CNPq), 
Coordena\c c\~{a}o de Aperfei\c coamento de 
Pessoal de N\'\i vel Superior (CAPES), Funda\c c\~{a}o de Amparo \`{a} 
Pesquisa do Estado do Rio de Janeiro (FAPERJ-BR), and Instituto Nacional 
de Ci\^encia e Tecnologia de Informa\c c\~ao Qu\^antica (INCT-CNPq).

\end{document}